\documentclass[floatfix,twocolumn,prl,showpacs,amsmath,superscriptaddress]{revtex4}

\usepackage{graphicx}
\usepackage{subfigure}
\usepackage{amssymb}
\usepackage{amscd}
\usepackage[usenames]{color}

\begin{document}
\title{The Power of Poincar\'e: Elucidating the Hidden Symmetries in Focal Conic Domains}
\author{Gareth P. Alexander}
\affiliation{Department of Physics \& Astronomy, University of Pennsylvania, 209 South 33rd Street, Philadelphia PA 19104, U.S.A.}
\author{Bryan Gin-ge Chen}
\affiliation{Department of Physics \& Astronomy, University of Pennsylvania, 209 South 33rd Street, Philadelphia PA 19104, U.S.A.}
\author{Elisabetta A. Matsumoto}
\affiliation{Department of Physics \& Astronomy, University of Pennsylvania, 209 South 33rd Street, Philadelphia PA 19104, U.S.A.}
\author{Randall D. Kamien}
\affiliation{Department of Physics \& Astronomy, University of Pennsylvania, 209 South 33rd Street, Philadelphia PA 19104, U.S.A.}
\affiliation{School of Mathematics, Institute for Advanced Study, Princeton, NJ 08540, U.S.A.}
\date{\today}

%--------------------------------------------------------------------------

\begin{abstract}
Focal conic domains are typically the ``smoking gun'' by which smectic liquid crystalline phases are identified.  The geometry of the equally-spaced smectic layers is highly generic but, at the same time, difficult to work with.  In this Letter we develop an approach to the study of focal sets in smectics which exploits a hidden Poincar\'e symmetry revealed only by viewing the smectic layers as projections from one-higher dimension. We use this perspective to shed light upon several classic focal conic textures, including the concentric cyclides of Dupin, polygonal textures and tilt-grain boundaries. \end{abstract}
\maketitle

%===================

In equally-spaced layered systems, such as smectics, idealized flat layers are rarely realized while textures permeated with focal conic domains are prevalent~\cite{friedel10,bouligand72,kleman09}.
They are beautiful and geometrically precise conic sections that are observed to occur in pairs of perfect confocal partners and with adjacent domains often exhibiting a more widespread level of geometric organization as in Friedel's law of corresponding cones~\cite{friedel10,kleman00}, the {\it treillis et r\'{e}seaux} expounded by Bouligand~\cite{bouligand72}, or Apollonian packings~\cite{bidaux73}. Since purely topological considerations are far too pliable to produce this level of geometric precision, focal conics must arise from a more rigid form of constraint; indeed, prior even to the knowledge of their molecular nature, it was realized that they are governed by the criterion of equal layer spacing~\cite{friedel10}. The condition of equal spacing, though an idealization, is a boon to the description of smectics, since it enables the form of the layers throughout the entire system to be determined uniquely~\cite{sethna82}, 
for instance from their focal sets. 
In this Letter we describe a hidden symmetry that underlies the structures of focal conics by viewing domains in $d$-dimensions, $\mathbb{R}^d$, as level sets of hypersurfaces in $d\!+\!1$-dimensional Minkowski space, $\mathbb{R}^{d,1}$. In this framework, the equal spacing constraint is equivalent to the condition that the hypersurfaces are lightlike, or {\it null}. Lorentz transformations, by their very construction, preserve the null condition; the level sets, however, are changed, giving rise to different domain geometries, which nonetheless arise from the same hypersurface.  

We quickly recall the symmetries and free energy of the smectic liquid crystal phase. The smectic phase is characterized by a one-dimensional density wave, $\rho({\bf x}) = \rho_0 + \rho_1\cos[2\pi\phi({\bf x})/a]$, where the level sets $\phi({\bf x})=na,\,n\in\mathbb{Z}$, with $a$ the layer spacing, define smectic layers with unit normal ${\bf N}=\nabla\phi/\vert\nabla\phi\vert$. The ground state consists of equally spaced, flat layers, and the free energy is
\begin{equation}
F =\frac{1}{2} \int d^d\!x\,\left\{ \frac{B}{4}\left[\left(\nabla\phi\right)^2-1\right]^2 + K \left(\nabla\cdot{\bf N}\right)^2\right\} ,
\end{equation}
where the first term, the compression, controls the spacing and the second the mean curvature of the lamellae. The energy is invariant under $\phi\rightarrow\phi + \hbox{constant}$, and ${\bf N}\rightarrow -{\bf N}$. The former symmetry represents a constant displacement of the smectic layers, but we see that $\phi\rightarrow\phi+a$ is merely a reparameterization of the density; likewise the nematic symmetry implies that $\nabla\phi\rightarrow -\nabla\phi$ results in precisely the same density wave, or equivalently $\phi\rightarrow-\phi$.  It follows that $\phi\in S^1/\mathbb{Z}_2$. In this Letter we will focus on configurations with vanishing compression.

Deferring the full symmetries of $\phi$ momentarily and instead working in the universal cover, $\mathbb{R}$, we recall that as the phase field associates a real number to every point ${\bf x}$ in the material, it is convenient to consider this as the coordinate of an extra dimension and view the smectic as a surface $\bigl({\bf x},\phi({\bf x})\bigr)$ in this larger space~\cite{chen09}. Equally spaced structures, which correspond to focal conic domains, play a privileged role. Examples of equally spaced smectics include the ground state $\phi = x$, say, where the surface is a plane, and the point defect $\phi = |{\bf x}|$, where the surface is a right circular cone. Writing this last example instead as $|{\bf x}|^2\!-\!\phi^2\!=\!0$ affords a more useful interpretation: the surface is the light cone of an event in Minkowski space (with $c=1$), with the value of the phase field viewed as a time-like direction. Importantly this correspondence is entirely general, as the condition of being null corresponds exactly to moving a unit distance in space for every unit change in $\phi$, thereby ensuring equal spacing. Moreover, since Lorentz transformations preserve null hypersurfaces {\it a fortiori}, focal conics also inherit this symmetry~\cite{footnote1}. . 
Defining, according to tradition, $\beta$ as the boost ``velocity'' and $\gamma=(1-\beta^2)^{-1/2}$, the Lorentz transformation is $x' = \gamma(x-\beta\phi)$, $y'=y$, and $\phi' = \gamma(\phi-\beta x)$~\cite{larmor97,einstein05}. We also bring the reader's attention to the fact that the symmetries relating different textures do not act on $\phi$ alone, but on a larger space~\cite{footnote2}. Just as a cube can cast a square, rectangular, or hexagonal shadow, so too do different ``projections'' of the same surface lead to different smectic textures, revealing an underlying universal structure: namely the same null hypersurface, just viewed by different ``observers.'' 

A general null hypersurface will exhibit points of ``singularity,'' like the apex of the light cone, through which multiple light rays pass. These points correspond to the focal sets of the smectic and provide the most convenient way of describing the texture.  It is precisely at these points that the projection of the surface normal is ill-defined, corresponding to disclinations and kinks in the nematic director, normal to the smectic layers. 
To construct the null hypersurface corresponding to any focal conic texture, it suffices to specify all of its focal sets. However, without prior knowledge of their form, the general case would seem a daunting task. Insight can be gained by first considering a class of null hypersurfaces whose focal sets form a dual pair. Not only does this immediately produce the precise shapes seen experimentally, but it also yields a natural generalization to multiple domain configurations that again captures the experimental features. 

\begin{figure}[t]
\centering
\includegraphics[width=87mm]{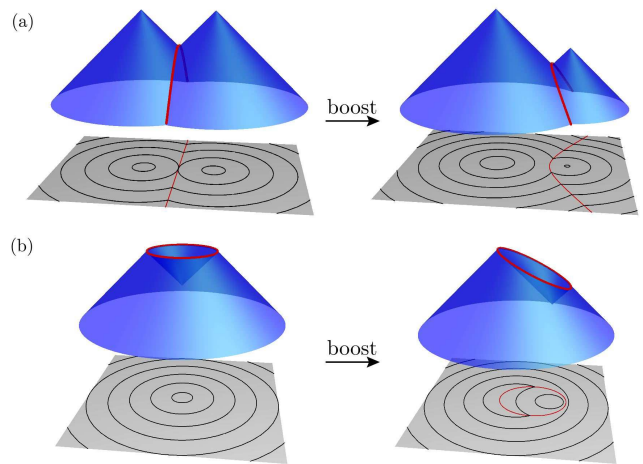}
\caption{(Color online) Achronal boundaries for (a) a pair of space-like and (b) time-like separated events, both in their rest frames (left) and in a general frame (right). The corresponding smectic textures are shown below each surface, with focal lines indicated in red.}
\label{fig:cones}
\end{figure}

Consider a pair of events in $\mathbb{R}^{2,1}$, which generate two-dimensional focal conics through the intersection of their light cones as shown in Fig.~\ref{fig:cones}. As is well known, a pair of events are either space-like, time-like, or null separated. Space-like separated events have coordinates $(\pm r,0,0)$ in their rest frame and the intersection of their light cones occurs on the hyperbola $y^2\!-\!\phi^2\!=\!-r^2$, in the $x\!=\!0$ plane. In a general frame, obtained by a boost along the $x$-direction, the two events lie at $(\pm \gamma r, 0, \mp \gamma\beta r)$, and the intersection of their light cones becomes $y^{\prime \,2}\!-\!(\phi^{\prime}/\gamma)^2\!=\!-r^2, \, x^{\prime}\!=\!-\!\beta\phi^{\prime}$. The smectic layers are given by equal time slices ($\phi^{\prime} \!=\! na$) of the null hypersurface formed by the two light cones. These are circles, or arcs of circles, concentric about $(x^{\prime},y^{\prime}) \!=\! (\pm \gamma r, 0)$ that are the vertical projections of the two events and the foci of one branch of the projected hyperbola $(x^{\prime}/\gamma\beta)^2\!-\!y^{\prime \,2}\!=\!r^2$, along which there are cusp singularities - precisely the focal lines seen in experiment. 

An identical analysis can be given for two time-like separated events, which in their rest frame have coordinates $(0,0,\pm r)$ leading to an intersection of their light cones on the circle $x^2\!+\!y^2\!=\!r^2$ in the $\phi\!=\!0$ plane. Boosting to a general frame as before, the two events take the coordinates $(\mp \gamma\beta r,0,\pm \gamma r)$, while their conjugate focal set becomes $(x^{\prime}/\gamma)^2\!+\!y^{\prime \,2}\!=\!r^2, \, \phi^{\prime} \!=\!-\! \beta x^{\prime}$, the equation of an ellipse. Again the projections of the two events coincide with the foci of the ellipse. The final case of null separation between the events is exceptional and we defer its treatment for the time being. 

In general, focal conics may be defined by their focal sets; for two space-like separated events these sets are $\Sigma = \{ x^2\!=\!r^2; y\!=\!\phi\!=\!0 \}, \overline{\Sigma} = \{ y^2\!-\!\phi^2\!=\!-r^2; x\!=\!0 \}$, while for two time-like separated events they are $\Sigma = \{ x^2\!+\!y^2\!=\!r^2; \phi\!=\!0 \}, \overline{\Sigma} = \{ -\phi^2\!=\!-r^2; x\!=\!y\!=\!0 \}$. In both cases these sets are mutually null separated and lie in orthogonal subspaces. In higher dimensions this simple structure persists and the focal sets take the form $\Sigma = \{ x^2\!+\!\vec{y}^2\!=\!r^2; \vec{z}\!=\!\phi\!=\!0 \}, \overline{\Sigma} = \{ \vec{z}^2\!-\phi^2\!=\!-r^2; x\!=\!\vec{y}\!=\!0 \}$, where $\vec{y}$ and $\vec{z}$ are $k$- and $(d\!-\!k\!-\!1)$-dimensional vectors, respectively, with $k\in \{0,1,..,d-1\}$.
This classification of simple null hypersurfaces corresponds with those described by Friedlander as being associated with progressive wave solutions of the wave equation~\cite{friedlander47}. As we shall see they serve as the building blocks for focal conic textures. The limiting cases $k\!=\!0$ and $k\!=\!d\!-\!1$ reduce to a pair of space-like and time-like separated events, respectively. In three dimensions there is only one other possibility, namely $k\!=\!1$ where the focal sets are both one-dimensional and correspond to a circle and an hyperbola lying in orthogonal subspaces. These are the elliptic-hyperbolic focal conic domains, which give smectic layers that are confocal cyclides of Dupin~\cite{friedel10,bouligand72,kleman09,cayley65,maxwell67}.  Our approach to the cyclides amounts to a Lorentzian phrasing of Maxwell's construction~\cite{maxwell67}.

A null hypersurface $S$ can be constructed from the focal sets as the union of all light rays connecting $e\in\Sigma$ to $\overline{e}\in\overline{\Sigma}$; $S=\{p\!=\!(e\!+\!\overline{e})/2\!+\!\xi(e\!-\!\overline{e})/2, \xi\in I_{e}^{\overline{e}}\}$ where $I_{e}^{\overline{e}}$ is a connected interval of $\mathbb{R}$ depdending on $e$ and $\overline{e}$. For the confocal hyperbola and ellipse, an explicit representation of $S$ follows from the form of the focal sets~\cite{friedlander47} $[(s+r)^2+z^2 - \phi^2][(s-r)^2+z^2-\phi^2]=0$,
where $s \!\equiv\! \sqrt{x^2+y^2}$ is the radius in cylindrical coordinates, $(s,\theta,z,\phi)$. This is precisely Cayley's quadric expression~\cite{cayley65} in its ``rest frame'', and thus reveals that the Dupin cyclides are a ``product of two cones.'' Note that there is no need to consider different cyclides or different types of elliptic-hyperbolic focal domains as they are all given by the same null hypersurface: one need only exploit Lorentz transformations and take different time slices. However, it is important to make the distinction between the propagation of light and smectic layers. In the former, the wavefronts can pass through each other, while in the latter they cannot. As shown in Fig. 1, for instance, this means that once the cones intersect the null hypersurface ends on this lower-dimensional cusp. In the parlance of general relativity, such surfaces are known as achronal boundaries~\cite{hawking73}. In general, for those light rays originating from a point $\overline{e}$ of $\overline{\Sigma}$ with $\phi\!<\!0$ we take $I_{e}^{\overline{e}}\!=\![-1,1]$, while for points with $\phi\!>\!0$ we take $I_{e}^{\overline{e}}\!=\![1,\infty)$. Note that $y$ and $z$ may be replaced with $\vec{y}$ and $\vec{z}$ without change to the foregoing discussion~\cite{kaneko86}. 

Elliptic-hyperbolic focal domains arise from a decomposition of Minkowski space into a pair of orthogonal subspaces, one space-like and one time-like, leading to focal sets that are spheres of square radius $\pm r^2$, one in each subspace. The only other construction of this kind is a decomposition of Minkowski space into a pair of orthogonal null subspaces. Denoting by $u_{\pm}$ the affine distances along the null directions $(1,0,0,\pm 1)$, we can take the null subspaces to be the $u_{+}y$ and $u_{+}z$ planes, separated by a distance $\sigma$ along the $u_{-}$ direction. One may swiftly verify that the sets $\Sigma \!=\! \{ 4\sigma u_{+}\!+\!y^2\!=\!0 ; \, u_{-}\!=\!\sigma/2, z\!=\!0 \}$ and $\overline{\Sigma} \!=\! \{ -4\sigma u_{+}\!+\!z^2\!=\!0 ; \, u_{-}\!=\!-\sigma/2, y\!=\!0 \}$ are null separated and thus serve as the focal sets for a null hypersurface, $S$, this time corresponding to a parabolic focal conic. Again these coincide with Friedlander's classification of progressing waves~\cite{friedlander47}.  

Although single focal domains correspond to null hypersurfaces in Friedlander's classification of progressive waves, this correspondence does not carry over to textures with more than one focal domain. Here we are tasked with the question of how to appropriately adjoin separate domains to form a larger structure, a task for which the Lorentzian viewpoint provides a convenient perspective. We provide two illustrative examples.

\begin{figure}[t]
\centering
\includegraphics[width=87mm]{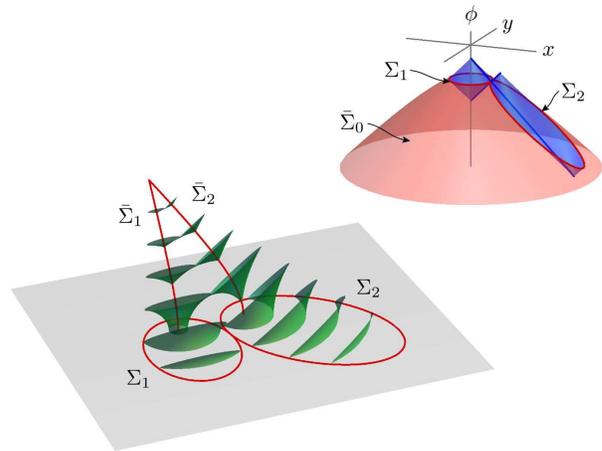}
\caption{(Color online) Two Dupin cyclide domains sharing a point of tangency in a trellis configuration. The cyclides fill conical regions extending from the point of intersection of their hyperbolae to their focal ellipse, while outside these regions the layers are continued by concentric spheres. In Minkowski space (inset, $z=0$ subspace) the ellipses lie on the surface of an hyperboloid. Tangency of ellipses corresponds to null separation of their foci, as is most easily seen in the rest frame of one of the ellipses.}
\label{fig:trellis}
\end{figure}

First, we consider the trellis configurations and Friedel's law of corresponding cones~\cite{friedel10}.  This is a collection of cyclidal domains organized so that their hyperbolae all intersect at a pair of points, as implied in Fig.~\ref{fig:trellis}. In Minkowski space these points of intersection represent a pair of space-like separated events, which, as we have seen, exhibit a conjugate focal set that is a hyperboloid (cf Fig.~\ref{fig:cones}). Friedel's laws are simply the statement that the ellipses of the individual cyclidal domains all lie on this hyperboloid.
Concretely, denoting the pair of events by $\Sigma_0 \!=\! \{ z^2\!=\!R^2; \, x\!=\!y\!=\!\phi\!=\!0 \}$, the conjugate hyperboloid is  $\overline{\Sigma}_0 \!=\! \{ x^2\!+\!y^2\!-\!\phi^2\!=\!-R^2; \, z\!=\!0 \}$. Observe that the focal sets $\Sigma_1 \!=\! \{ x^2\!+\!y^2\!=\!r^2; \phi\!=\!-\sqrt{r^2\!+\!R^2}, z\!=\!0 \}, \overline{\Sigma}_1 \!=\! \{ z^2\!-\!(\phi\!+\!\sqrt{r^2\!+\!R^2})^2\!=\!-r^2; x\!=\!y\!=\!0 \}$, are cyclidal domains, exhibiting the desired nesting $\Sigma_1 \!\subset\! \overline{\Sigma}_0,\, \overline{\Sigma}_1 \!\supset\! \Sigma_0$, so that the focal ellipse of this domain lies on the surface of the hyperboloid $\overline{\Sigma}_0$, while the focal hyperbola passes through the original pair of events, $\Sigma_0$. Importantly, this is preserved by Lorentz transformations and since these act transitively on $\overline{\Sigma}_0$ they in fact generate all focal sets $(\Sigma_1,\overline{\Sigma}_1)$ with this property. 

The null hypersurface for the composite texture is formed by taking the surface for the concentric sphere domain and omitting those light rays connecting $\Sigma_0$ to the part of $\overline{\Sigma}_0$ inside the circle $\Sigma_1$. These are then replaced by light rays connecting $\Sigma_1$ to $\overline{\Sigma}_1$. In this way the excised region of the concentric sphere domain is filled with the cyclides of Dupin~\cite{sethna82} and because the interface consists of the same light rays, both the smectic layers and the layer normal are continuous across the join. Of course, this construction can be repeated for a collection of focal domains $\{ \Sigma_i,\overline{\Sigma}_i \}_{i=1}^{N}$ to yield a trellis structure~\cite{bouligand72}. Tangency of adjacent ellipses corresponds to their foci being null separated with the light rays connecting them passing through their common point, as is readily apparent when viewed from one of their rest frames (Fig.~\ref{fig:trellis}). When projected into $\mathbb{R}^d$ this is Friedel's observation that straight lines can be drawn between the foci of touching domains that pass through their point of tangency~\cite{friedel10}. 

\begin{figure}[t]
\centering
\includegraphics[width=65mm]{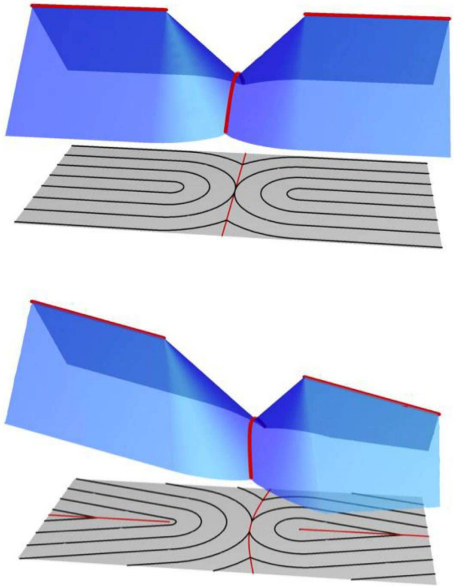}
\caption{(Color online) Two-dimensional slice ($y=0$) of a tilt grain boundary configuration, shown both in real space and in Minkowski space. Note the ``ridge'' at which there is a change in orientation of the ground state: under a boost (bottom) this ridge becomes an additional focal set.}
\label{fig:tilt}
\end{figure}

Tilt-grain-boundaries~\cite{bidaux73,kleman00,blanc00} involve a similar replacement of a region of one type of texture with that of another, and again this can be conveniently achieved by working in the appropriate rest frame. In this instance, one removes a cylindrical region of the ground state and fills it with the inner part of a cyclide domain. This cyclide region is described by those light rays connecting the elliptical focal set $\Sigma \!=\! \{ x^2\!+\!y^2\!=\!r^2; \, z\!=\!\phi\!=\!0 \}$ to the branch of the hyperbola $\overline{\Sigma} \!=\! \{ z^2\!-\!\phi^2\!=\!-r^2; \, x\!=\!y\!=\!0 \}$ with $\phi \!<\! 0$. Moving out along the hyperbola towards $z \rightarrow \pm \infty$ the light rays from any point of the ellipse asymptote onto the directions $(0,0,\pm 1,-1)$ corresponding to the equivalent ground states $\phi\!=\!\mp z$ seen outside of the cyclides, Fig.~\ref{fig:tilt}. Importantly, although these are equivalent they are not the same ground state, but differ by a change in orientation occuring at the ellipse $\Sigma$, which here has the appearance of a $+\tfrac{1}{2}$ disclination loop. A boost along the $x$-direction rotates the asymptotic directions to $(\gamma\beta,0,\pm 1,-\gamma)$, {\it i.e.}, the cyclide region now connects the ground states $\phi^{\prime} \!=\!-\! \beta x^{\prime} \mp \gamma^{-1} z^{\prime}$, rotated relative to each other by $2\arcsin (\beta)$. Since the boosted ground states are no longer equivalent, they no longer join smoothly on the plane, $z^{\prime}\!=\!0,\phi^{\prime}\!=\!-\!\beta x^{\prime}$, but rather form a plane of cusps (Fig.~\ref{fig:tilt}). The tilt-grain-boundary construction provides one example of this generic behavior, that additional focal sets are produced when a non-orientable texture is subjected to a Lorentz transformation.  Finally let us remark that, although we have considered only one cyclide region, it is clear that any number can be accommodated by tiling the $z\!=\!\phi\!=\!0$ plane of the rest frame with circles, each circle being the elliptical focal set of a cyclide domain.  

This Letter has demonstrated the connection between focal conics in $\mathbb{R}^d$ and null hypersurfaces in $\mathbb{R}^{d,1}$. Specifically, in three dimensions we have shown that simple textures with codimension 2 focal sets arise from intersections of cones and planes. The addition of the extra dimension clarifies the action of the isometries, just as an extra dimension reveals the simplicity of the M\"obius Transformations of the plane in terms of the symmetries of the Riemann sphere~\cite{Arnold}. Further work will elucidate other smectic structures, such as Apollonian packings and oily-streak textures and will demonstrate that energetic calculations are natural in rest-frame coordinates~\cite{LongPaper}. Our approach suggests that methods of general relativity will be fruitfully used for the study of defects on curved surfaces~\cite{curved}. We hope that the structure of defects presented here along with the results of~\cite{chen09} set the framework for a combined theory of disclinations, focal sets, and dislocations.

It is a pleasure to acknowledge discussions with G.W. Gibbons, H. Karcher, R. Kusner, and J. Maldacena. This work was supported by DMR05-47230 and gifts from H.H. Coburn and L.J. Bernstein.

%=========================


\begin{thebibliography}{99}
\bibitem{footnote1} Note that Lorentz covariance is a symmetry of the smectic texture not shared by the free energy functional.
\bibitem{footnote2} The formal procedure is a sequence of maps:\hfill\break
\centerline{$
\mathbb{R}^d \; \xrightarrow{\Phi} \; \mathbb{R}^{d,1} \; \xrightarrow{\pi\circ\Lambda_{\beta}} \; \mathbb{R}^d\times\mathbb{R} \; \xrightarrow{(1,\tau)} \; \mathbb{R}^d\times\bigl( S^1/\mathbb{Z}_2 \bigr) 
$}\break
The first generates a null hypersurface, upon which act Lorentz transformations $\Lambda_{\beta}$. 
To then construct a smectic texture consistent with the symmetries of
a density wave
we fix a frame to return from Minkowski to Euclidean space using a map
$\pi$ that takes $(x, \phi)\in\mathbb{R}^{d,1}$ to
$(x,\phi)\in\mathbb{R}^d\times\mathbb{R}$,
and subsequently project the former time coordinate $\phi$ to a phase
in $S^1/\mathbb{Z}_2$ .
\end{thebibliography}
\end{document}